\documentclass[aps,prb,twocolumn,,amsmath,amssymb,floatfix,nofootinbib,
 superscriptaddress,showpacs]{revtex4}
\usepackage{graphicx}      
\usepackage{bm}            
\usepackage{bbm}
\usepackage[dvips]{epsfig,color}

\def \beq{\begin{equation}}
\def \eeq{\end{equation}}
\def \beqarr{\begin{eqnarray}}
\def \eeqarr{\end{eqnarray}}
\def \bspt{\begin{split}}
\def \espt{\end{split}}
\def \bef{\begin{figure}}
\def \enf{\end{figure}}

\def \bpm{\begin{pmatrix}}
\def \epm{\end{pmatrix}}

\newcommand{\abs}[1]{\lvert#1\rvert}

\begin{document}

\title{Edge States at the Interface between Monolayer and Bilayer Graphene}

\author{Zi-Xiang Hu}
\affiliation{Department of Physics, ChongQing University, ChongQing, 440004, China}
\affiliation{Department of Electrical Engineering, Princeton University, Princeton, New Jersey
08544, USA}

\author{Wenxin Ding}
\affiliation{NHMFL and Department of Physics, Florida State
University, Tallahassee, Florida 32306, USA}

\date{\today}
\begin{abstract}
The electronic property of monolayer-bilayer hybrid graphene with a
zigzag interface is studied by both the Dirac equation and numerical calculation. There
are two types of zigzag interface stacks. The dispersion and local density of
states behave quit differently along the interface at the Fermi energy due to the different locations of the edge state.
We hope our study can give some insights in the understanding of the transport and STM experiments.
\end{abstract}
\maketitle

\section{Introduction}
In recent years, the experimental accessibility of the single and multi
layered graphene
samples~\cite{Novoselov,Yzhang,Bunch,Berger,Novoselovpnas}
has attracted considerable theoretical and experimental attention due
to its unusual electronic structure described by the Dirac equation,
namely electrons in monolayer graphene have linear dispersion thus
behave like massless Dirac fermions at the corners of the Brillouin
zone (BZ)~\cite{Wallace}.  In the presence of magnetic field
perpendicular to the graphene plane, the system shows anomalous
integer quantum hall
effect~\cite{Novoselovnature,Yzhangnature,Novoselovscience,Abanin} (IQHE)
which is different from that of the conventional two-dimensional
electron system in semiconductor heterostructures.
The Hall conductivity in the IQHE of graphene shows half integer
plateaus instead of integer ones in the IQHE of a normal
semiconductor due to the different degeneracy at $N=0$ Landau level.

Edge states in graphene have been the focus of much theoretical study
because of the important role they play in transport.~\cite{castrormp}
It is well known that there are two basic types of edges in graphene,
namely, the armchair and zigzag edges. Some theoretical
work~\cite{stein1987,Tanaka,Fujita,Fujitapsj,Nakada,zhenghx,brey,
  Peres, kohmoto, Ryu, sasaki, sasaki07, castro, Niimi, kobayashi} on
the electronic structure
of finite-sized systems, either as molecules or as one-dimensional
systems, has shown that graphene with zigzag edge has localized edge
states at the Fermi energy, but those with armchair edges do not
have such state. Therefore, the zigzag edge states is essential to the
transport properties in graphene since the localized edge state has
contribution to the conductivity. However, most of the work were focused on
the uniform monolayer and bilayer graphene. On the other hand the
hybrid edge structure composed of partial monolayer and partial
bilayer graphene , which is quit general in reality, has received
not so much attention.  Experimentally,~\cite{yliu} the anomalous
quantum oscillations in magnetoconductance was observed due to the
peculiar physics along the interface. Some of the transport properties
and the presence of interface Landau levels was explored within an
effective-mass approximation.~\cite{Koshino, Takanishi}

In this work\cite{wenxin}, we study the electronic properties of the
hybrid interface systematically via both tight binding model and
its effective theory in the continuum limit - the Dirac
equation. The
edge states in graphene can be studied experimentally by using a local
probe such as scanning tunneling microscopy
(STM). The STM experiments measure the differential conductance which
is proportional to the density of states. Thus we study the local
density of states (LDOS) in different hybridized zigzag edge graphene. It is defined as:
\beq
N(r, eV) = |\Psi_\alpha(r)|^2 \delta(eV - E_\alpha),
\eeq
where $\Psi_\alpha(r)$ is the eigenfunction with energy
$E_\alpha$. The LDOS shows the strength of the local electronic
density which is related to the strength of the signal in STM
experimental data. Therefore, the LDOS can show us the signals of the
different edge state as in our previous work.~\cite{hu} In this
paper, our study of the step edge graphene shows that there are two
different zigzag step edge, and there always exists
zero energy states localized near the zigzag step edge but the
distribution of LDOS of these edge states strongly depends on the
details of how the edge stacks together. We find the energy dispersion around the Dirac cones also 
presents different characters for different edge arrangements either within or without a magnetic field.

This paper is arranged as follows: In section II, we set up the model
Hamiltonian in different geometries.  The zero energy solutions in
zero field and in magnetic field are obtained by solving the Dirac
equations and numerical diagonalization is implemented in a finite
system in section III and IV, and some discussions and conclusions are
in the section V.

\section{Model and Different Geometries}
\begin{figure}
\includegraphics[width=4cm]{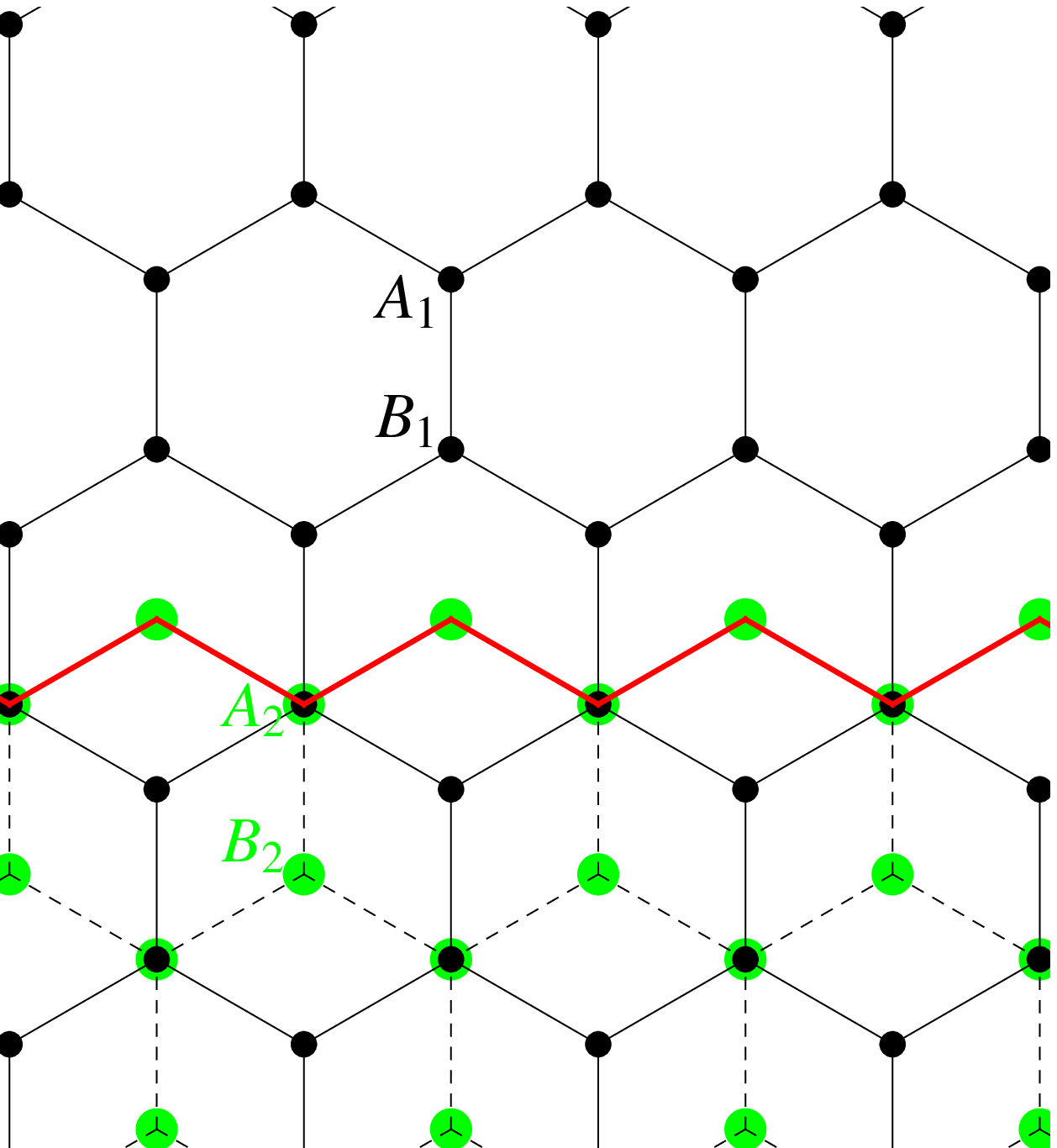}
\includegraphics[width=4cm]{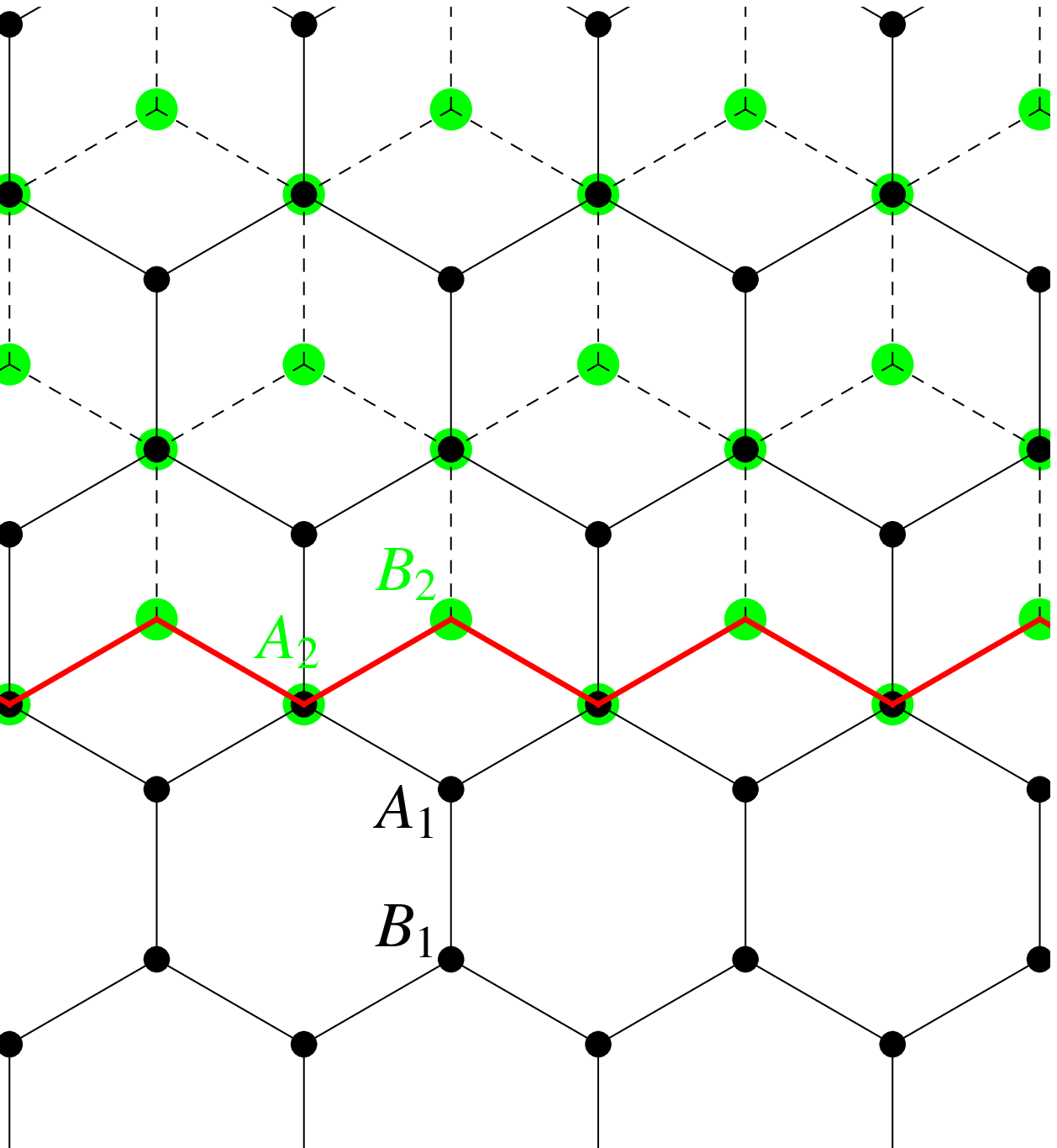}
\caption{\label{fig:ab_stacking}(Color online)The schematic pictures for two kinds
  of monolayer-bilayer interface.  The atoms of the extended bottom
  layer (layer 1) are indicated by black dots while the green dots
  represent the top layer which terminates at the interface. In the left plot, the top layer
 is ended with $B_2$ sites ({\it l.e.b}) while on the right it is
 ended with $A_2$ sites ({\it h.e.b}).}
\end{figure}

We consider the bilayer graphene with $AB$ stacking, as shown in
Fig. \ref{fig:ab_stacking}; its tight-binding Hamiltonian can be written as:
\beqarr
H = &-&t \sum_{i=1}^2 \sum_{m,n} a^\dagger_{i;m,n} (b_{i;m,n} +
b_{i;m-1,n} + b_{i;m,n-1}) \nonumber\\
&-& t_{\perp} \sum_{m,n} a^\dagger_{2;m,n}
b_{1;m,n} + h.c.,
\eeqarr
where $a_{i;m,n}$ ($b_{i;m,n}$) is the annihilation operator at
position ($m,\ n$) in sublattice $A_i$ ($B_i$), and $i=1,\ 2$, indicating
the two layers.
The first term is the Hamiltonian within each layer,
and the second term describes the interlayer coupling
in which we only consider the hopping between the two atoms stacked
right on top of each other. Let us label the bottom
(extended) layer as layer $1$, the half plane upper layer as layer
$2$. In this work we will only consider the $A-B$ (Bernal)
stacking. One can expand the effective Hamiltonian near the two Dirac
points $K$ and $K'$ which are time reversal symmetric partners. In
momentum space, the hamiltonian near $K$ can be written as:
\beq
H = \sum_k \Psi^\dagger_k \cdot H_k \cdot \Psi_k,
\eeq
where
\beq \label{hambilayer}
H_k = \begin{pmatrix} 0 & v_F k & 0 & 0 \\ v_F k^* & 0 & t_{\perp} &
  0 \\ 0 & t_\perp & 0 & v_F k \\ 0 & 0 & v_F k^* & 0 \end{pmatrix}
= v_F \begin{pmatrix} 0 & k & 0 & 0 \\ k^* & 0 & \gamma &
  0 \\ 0 & \gamma & 0 & k \\ 0 & 0 & k^* & 0 \end{pmatrix},
\eeq
in which $k = k_x + i k_y$, $\gamma = t_\perp / v_F$, and $\Psi_k =
(a_{1;k},b_{1;k},a_{2;k},b_{2;k})$. For the other Dirac point, as
stated before, $H_{K'} = H_{K}^*$. Here we only consider the zigzag type interface(or edge) to explore the localized edge
state. Without magnetic field and the interface,  it is sufficient
to discuss just one Dirac cone in the continuum model due to the
symmetry. But with the interface breaking the inversion symmetry and
the magnetic field breaking the time reversal symmetry, the two Dirac
points are not equal to each other, and both must be studied.

According the lattice orientation we
adopt which is shown in Fig.~\ref{fig:ab_stacking}, the zigzag
interface is along the $x$ direction. For simplicity, we consider an
infinite stripe along the $x$ direction, therefore the system has
translational symmetry along the $x$ direction thus $k_x$ remains a good
quantum number. We then do Fourier transformation in the $x$ direction
and reduces the 2D problem to 1D. There are actually TWO distinct
geometries which are physically different.  i) The outmost sites
of the upper layer are the $B_2$ sites  which do not stack directly on
the lower layer atoms as showed in
Fig.~\ref{fig:ab_stacking}(left). The $B_2$ sites are the low energy
degrees of freedom (along with $A_1$) which are kept if one further
considers an $2\times2$ effective theory on energy scale $\epsilon \ll
t_{\bot}$. We label it the low energy sites boundary ({\it  l.e.b.});
ii) the $A_2$ sites are the outmost sites on the upper layer as showed
in Fig.~\ref{fig:ab_stacking}(right). The $A_2$ sites, together with
the $B_1$ sites which they stack right on top of, form the dimer
sites. In the $2\times2$ low energy effective theory the wavefunction
have almost zero weight on those dimes sites when $\epsilon \ll
t_{\bot}$. As a result these dimers are ignored in this limit. In
other words, they are occupied considerably only at high energy
(comparing to $t_{\bot}$). Therefore, we shall
address such interface as the high energy sites boundary ({\it h.e.b.}).

\section{Interface properties in Zero Field}
For a semi-infinite sheet, it is well known that the existence of the zero energy
edge modes in both monolayer and bilayer graphene.
Presumably, such modes are also expected at the interface between
them. Let us consider the following geometry: a
half plane of monolayer graphene and a half plane of bilayer graphene
joined along the zigzag edge; and look for solution(s) with zero
eigenenergy by using the Dirac equation.

Before we present the results, let us discuss the boundary conditions
at the interface firstly as in Ref.~\onlinecite{Nilsson}.  Let
\beq \Psi_{\text{mono}}(x,y) = \binom{\psi_A(x,y)}{\psi_B(x,y)} \eeq  and
\beq \Psi_{\text{bi}}(x,y) = \begin{pmatrix} \psi_{A1}(x,y) \\  \psi_{B1}(x,y)
  \\  \psi_{A2}(x,y) \\  \psi_{B2}(x,y)\end{pmatrix} \eeq
be the wavefunctions at Dirac point in the monolayer and bilayer respectively. Suppose the
interface locate at $y = 0$, the boundary condition for monolayer is then
straightforward: both components of the wavefunction must be
continuous. For the upper layer, it terminates at $y=0$ and therefore
satisfies the open boundary condition. Note that the term `terminates'
indicates the last row of lattice sites are the high/low energy sites,
however, the boundary condition is not the wavefunctions being zero
on these sites. They should be the wavefunctions on sites one unit
cell `outside' the boundary being zero. To sum up, we have:
\beqarr
&&\psi_A(x,0) = \psi_{A1}(x,0),\qquad \psi_B(x,0) = \psi_{B1}(x,0), \nonumber\\
&&\begin{cases} \psi_{A2}(x,0) = 0 & l.e.b. \\ \psi_{B2}(x,0) = 0 &
  h.e.b. \end{cases}.
\eeqarr

In the monolayer region with zigzag interface,
we do the substitution
$k\rightarrow k_x+\partial _y$ in the Dirac hamiltonian $H_{\text{mono}} =v_F \left( \begin{array}{cc}
 0 & k \\
 k^* & 0
\end{array}
\right)$.~\cite{castrormp} The zero energy solution is
\beq
\left\{
\begin{array}{c}
  \phi_A(y)=e^{y k_x} C_2 \\
 \phi_B(y)=e^{-y k_x} C_1
\end{array}
\right.
\eeq
In analogy, we do the same substitution in the bilayer Hamiltonian (Eq. ~\ref{hambilayer}) and obtain its zero energy solution:
\beq
\begin{cases}
 \phi_{B1}(y)=e^{-y k_x} A_1 \\
 \phi_{B2}(y)=-e^{-y k_x} y \gamma  A_1+e^{-y k_x} A_2 \\
 \phi_{A1}(y)=e^{y k_x} A_3 + e^{y k_x} y \gamma  A_4\\
 \phi_{A2}(y)=e^{y k_x} A_4
\end{cases}.
\eeq

Applying the requirement that the wavefunctions remain finite at
infinity, for the {\it l.e.b.} interface (so
$\phi_{A2}(y=0)=0$), one easily find that nonzero solutions only exist
for $k_x > 0$. The solution is
\beq
C_1 = C_2 = A_1 = A_3 = A_4 = 0, \qquad A_2 = \text{const}.
\eeq
For the {\it h.e.b.} interface, one finds that for $k_x > 0$
\beq
C_1 = C_2 = A_1 = A_2 = A_3 = 0, \qquad A_4 = \text{const}.
\eeq
For the other Dirac cone, the solutions remain the same but only exist
for $k_x < 0$. The only nonzero constant is to be determined by
normalization. Both results are in agreement with
tight-binding analysis\cite{castro-2008-84}.

Even though the zero energy states exist for both types of interface,
there is an important difference between them. In the $l.e.b.$ case, the
wavefunction is only none-zero in the upper layer, which is trivial
as such mode is expected when a graphene sheet is terminated at a
zigzag edge. However, the $h.e.b.$ is less trivial. The wavefunction
also lives on the extended layer at the interface where no cut is
present. We interpret this as, when the dimer sites are the boundary,
the interlayer coupling  $t_\bot$ imposes an energy cost for electrons
to go through the interface in the extended layer which can be
considered as an effective potential barrier. The potential barrier can localize the electron states along the interface.

\begin{figure}
\includegraphics[width=8cm]{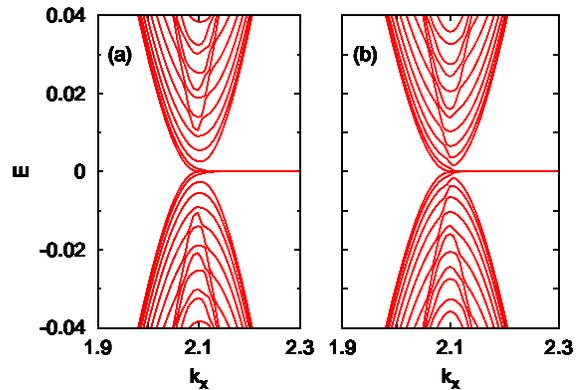}
  \caption{\label{fig:disp_step}The dispersion around Dirac cone for
    the step graphene with (a)l.e.b and (b)h.e.b respectively.}
\end{figure}

\begin{figure}
\includegraphics[width=4.cm]{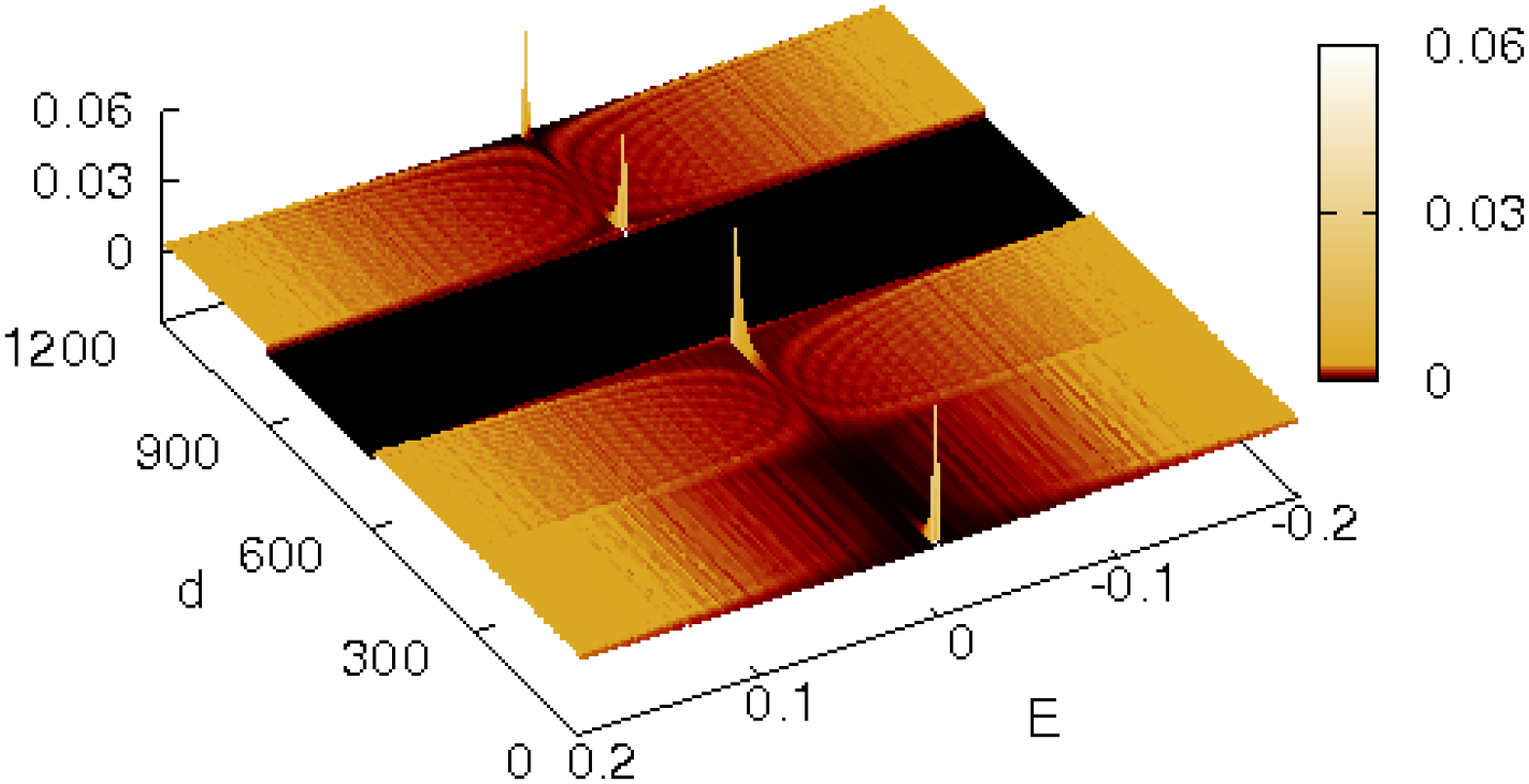}
\includegraphics[width=4.cm]{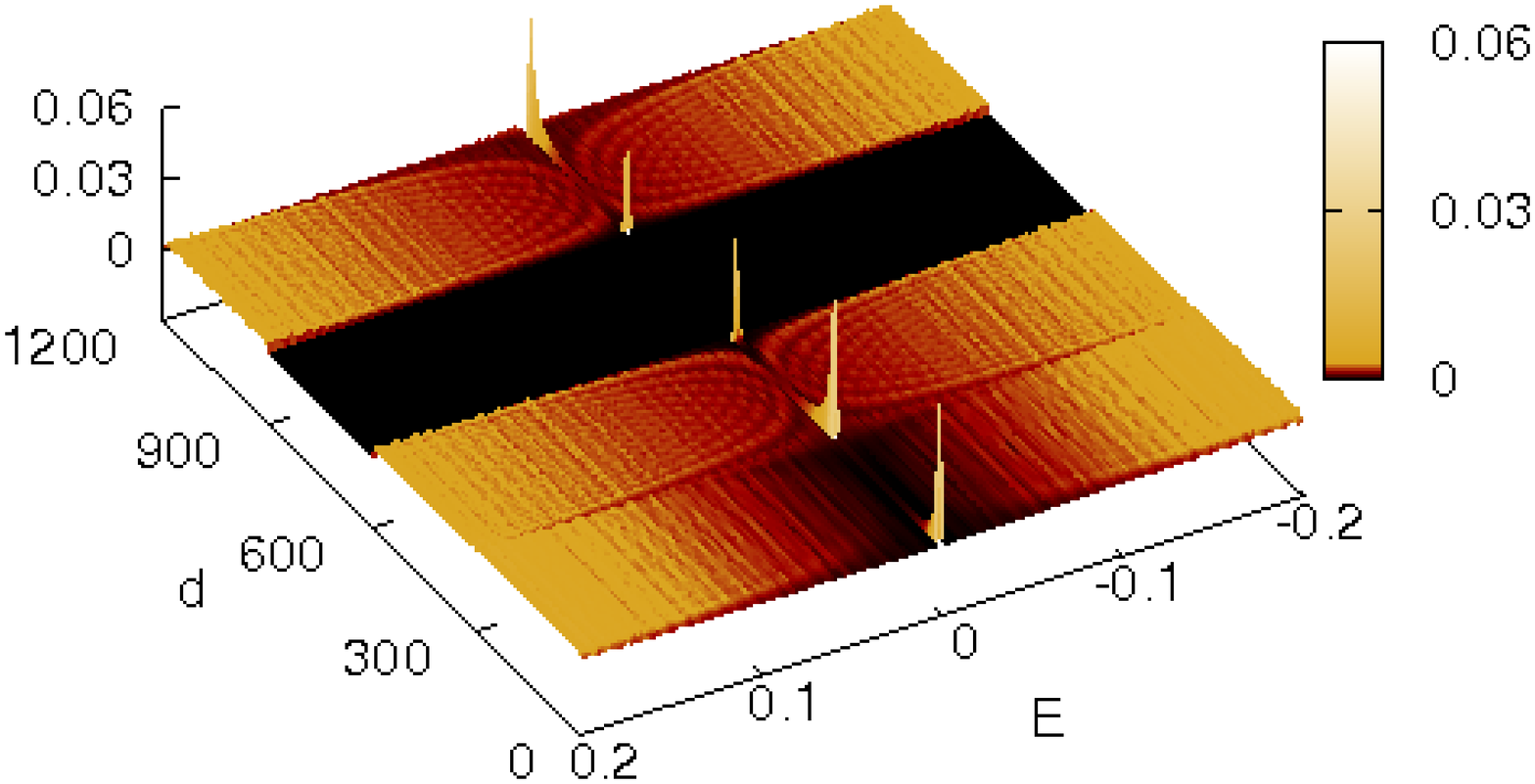}
\caption{\label{nofieldldos}(color online) The LDOS of the
step bilayer graphene with l.e.b.(left) and h.e.b.(right) respectively.
The width of of monolayer $L = 600$. In plot, the region $d \in
[0:600)$ stands for the lower extended layer and $d \in [600,1200)$ is
    the upper layer. }
\end{figure}

We numerically diagonalize a system with a finite width up
to 600 unit cells in the $y$ direction. The intralayer hopping strength
$t$ is set to identity and the interlayer hopping strength
$t_\perp = 0.2t$.  The dispersion relation is shown in Fig.~\ref{fig:disp_step}.
Compare the dispersions in different geometries, the {\it h.e.b.} edge 
has an obviously stronger level anti-crossing feature than the other. The
reason is that for {\it l.e.b.} edge, the zero energy edge state just
locates on the bilayer graphene which has quadratic dispersion and
has nothing to do with the monolayer part with a linear
dispersion. However, in the case of {\it h.e.b.}, the zero energy edge
state also has component on the monolayer graphene at the interface,
the linear dispersion part for monolayer should also connect to the edge state which
induces more energy level anticrossing around the Dirac points.
Fig.\ref{nofieldldos} shows the  LDOS for the two kinds of step
graphene. We label the coorindinate of the extended layer by $d \in [0,600)$ and the 
$d \in [900, 1200)$ for upper layer. Therefore, the interface locates at
$d=300$ and $d=900$ for layer 1 and 2 respectively.
One can notice that the localized edge state shows as a peak at zero energy
 along the interface. The prominent difference is that the peak just appears on the top layer of bilayer part in
the {\it l.e.b.} case; but for the {\it h.e.b.} case, the zero energy
peak appears on both two layers although still
locates at the bilayer side. The numerical results are in
agreement with the analysis by the Dirac equations
and we conclude that the two kinds of edge arrangements should have
different consequences in experiments since the different distributions of the zero mode. Here we notice that the other
peaks in LDOS at the end of the finite system is the signal of the
general monolayer or bilayer zigzag edge graphene as discussed in many others
work.~\cite{stein1987,Tanaka,Fujita,Fujitapsj,Nakada,zhenghx,brey, Peres, kohmoto, Ryu, sasaki, sasaki07, castro, Niimi, kobayashi}

\section{Interface properties in Magnetic Fields}

In the presence of magnetic field, by minimal coupling the Dirac
equation should be modified by doing the substitution ${\bf k}
\rightarrow {\bf k} + \frac{e {\bf A}}{c}$. Assume the magnetic field
is along the direction of perpendicular to the plane $B =  B
\hat{z},\ B>0$. We adopt the Landau gauge here $\vec{A} = (A_x,
A_y) = (- y B, 0)$ due to the translational invariant along $x$
direction. Thus the Hamiltonian of monolayer becomes:
\beq
H_{\text{mono}} = \begin{pmatrix} 0 &  k_x + \partial_y -
  \frac{eBy}{c}\\ k_x - \partial_y - \frac{eBy}{c} &
  0\end{pmatrix}.
\eeq
 and its zero energy solution is
\beq
\left\{
\begin{array}{c}
 \phi_A(y)=C_1 e^{k_x y - \frac{eB}{2c}y^2}  \\
 \phi_B(y)=C_2 e^{-k_x y+ \frac{eB}{2c}y^2}
\end{array}
\right.
\eeq
Similarly, the zero energy solution for bilayer graphene becomes:
\beq
\begin{cases}
 \phi_{A1}(y)= (A_3  \gamma y + A_2 )e^{k_x y - \frac{eB}{2c}y^2 }\\
 \phi_{B1}(y)= A_1 e^{- k_x y + \frac{eB}{2c} y^2}\\
 \phi_{A2}(y)=A_3 e^{k_x y-\frac{eB}{2c} y^2}\\
 \phi_{B2}(y)=(- A_1  \gamma y + A_4 )e^{- k_x y + \frac{eB}{2c}y^2 }
\end{cases}.
\eeq
Apply the same boundary conditions as in the zero field case, we will
have the solutions. For $l.e.b.$, one gets
\beq
C_2 = A_1 = A_3 = A_4 =0, \qquad C_1 = A_2 = \text{const}.
\eeq
For $h.e.b.$ the solution is
\beq
C_2 = A_1 = A_4 =0,  C_1 = A_2 = \text{const}_1, A_3 = \text{const}_2.
\eeq
The constants are to be determined by normalization conditions. One
immediately notices that the wavefunction lives only on the $A$ sublattice.
We should note that on the other Dirac point, the solutions remain the
same form, but resides on the $B$ sublattice. Another important
feature of the solution is that for $h.e.b.$ we actually have TWO
independent solutions here.

\subsection{Dispersion Relation}
The general problem in magnetic field can be written as
\beq
\frac{1}{\sqrt{2}}\left(
\begin{array}{cc}
  0  &  \partial _{\text{\textit{$\xi $}}}+\xi   \\
  -\partial _{\text{\textit{$\xi $}}}+\xi  &  0
\end{array} \right) \Psi_{\text{mono}} =\epsilon  \Psi_{\text{mono}},
\eeq
for the monolayer
while  that for the bilayer is
\beq
\frac{1}{\sqrt{2}} \left( \begin{array}{cccc}
  0  &  \partial _{\xi}+\xi   & 0 & 0 \\
  -\partial _{\xi}+\xi  &  0 & \tilde{\gamma} & 0
  \\
0 & \tilde{\gamma} & 0 &  \partial _{\xi}+\xi  \\
0 & 0 &  -\partial _{\xi}+\xi  & 0 \\
\end{array} \right)  \Psi_{\text{bi}} =\epsilon  \Psi_{\text{bi}}
\eeq
where $
\xi =\frac{y}{l_B}-l_Bk_x
\text{, }l_B=\sqrt{\frac{c}{e B}}, \epsilon =\frac{E}{\omega _c},
\omega _c=\sqrt{2}\frac{v_F}{l_B}$, $\tilde{\gamma} =
\frac{\gamma}{v_F \omega_c}$.
It is known that in the bulk, the solution to the monolayer is
\beq
\Psi_{\text{mono}} = \begin{pmatrix} \frac{1}{\Gamma(\epsilon^2 )}
  D_{\epsilon^2-1} (\sqrt{2} \xi)
  \\ \pm \frac{1}{\Gamma(\epsilon^2 + 1)} D_{\epsilon^2} (\sqrt{2}
  \xi) \end{pmatrix}  = \begin{pmatrix} \psi_{\epsilon^2-1}(\xi)
  \\ \pm \psi_{\epsilon^2}(\xi) \end{pmatrix} ,
\eeq
where the $\epsilon = \pm \sqrt{N}$, $N=0, 1, 2 \dots$, and $D_\nu$'s
are the parabolic cylinder functions,  which combined with the factor
$1/\Gamma(v+1)$ give us the eigen wavefunctions of a harmonic
oscillator $\psi_{\nu}(\xi)$. The bulk solution to the bilayer
Hamiltonian can be written in a similar way:
\beq
\Phi _{\text{bi}} =\left(
\begin{array}{c}
 \frac{\epsilon  \left(\epsilon ^2-(j+1)-\tilde{\gamma}
   ^2\right)}{\tilde{\gamma}  \sqrt{j (j+1)}}  \psi _{j-1} \\
 \frac{\epsilon ^2-(j+1)}{\tilde{\gamma}  \sqrt{j+1}} \psi _j \\
 \frac{\epsilon }{\sqrt{j+1}} \psi _j \\
 \psi _{j+1}
\end{array}
\right),
\eeq
where $\epsilon =\pm \sqrt{\frac{1+\tilde{\gamma} ^2+2 j\pm
    \sqrt{\left(1+\tilde{\gamma} ^2\right)^2+4\tilde{\gamma} ^2 j}}{2}}$ ($j = 0, 1,
  2, \dots$) is the eigen energy.

However, for the interface problem, we need solutions on the
half-plane. In this case, we let the bilayer live on the $y>0$ side,
so the monolayer is on the $y<0$ part. Therefore, for the bilayers,
the solutions on $(0, \infty)$ take on the same form as in the bulk,
but $j$'s are no longer required to be integers. Instead, we now
should replace $j$'s by
\beq
j_{1,2} = \epsilon ^2-\frac{1}{2}\pm \sqrt{\epsilon ^2 \tilde{\gamma}
  ^2+\frac{1}{4}},
\eeq
and $\epsilon$ now varies continuously. For the monolayer, the
solution on $(-\infty, 0)$ can be chosen as
\beq
\Psi_{\text{mono}} = \begin{pmatrix} \psi_{\epsilon^2-1}(-\xi)
  \\ \mp \psi_{\epsilon^2}(-\xi) \end{pmatrix} .
\eeq
With the above solutions, and combined with the boundary conditions
that are already discussed in the zero field cases,  we obtain the
transcendental equations that dictates the dispersion relations. For
the $l.e.b.$ interface,
\beqarr
&&\hspace{-0.5cm}(\epsilon ^2-(j_1+1)) D_{j_1-1}(-\sqrt{2}k_x) (D_{\epsilon ^2}(\sqrt{2} k_x)
D_{j_2}(-\sqrt{2} k_x) \nonumber\\
&& + D_{\epsilon ^2-1}(\sqrt{2}k_x) D_{j_2+1}(-\sqrt{2} k_x)) \nonumber\\
&&\hspace{-0.5cm}= (\epsilon^2-(j_2+1)) D_{j_2-1}(-\sqrt{2} k_x)
(D_{\epsilon ^2}(\sqrt{2} k_x) D_{j_1}(-\sqrt{2}k_x) \nonumber\\
&& + D_{\epsilon ^2-1}(\sqrt{2} k_x)D_{j_1+1}(-\sqrt{2} k_x)),
\eeqarr
where we have set $l_B = 1$. For the $h.e.b.$, the dispersion equation
is
\beqarr
&&\hspace{-0.5cm} (\epsilon ^2-(j_1+1)-\gamma ^2)
D_{j_1}(-\sqrt{2} k_x) (D_{\epsilon ^2}(\sqrt{2}
k_x) D_{j_2}(-\sqrt{2} k_x) \nonumber\\
&& +D_{\epsilon  ^2-1}(\sqrt{2} k_x) D_{j_2+1}(-\sqrt{2} k_x))  \nonumber\\
&&\hspace{-0.5cm} =(\epsilon ^2-(j_2+1)-\gamma ^2)D_{j_2}(-\sqrt{2} k_x) (D_{\epsilon ^2}(\sqrt{2}
k_x) D_{j_1}(-\sqrt{2} k_x)  \nonumber\\
&&+D_{\epsilon  ^2-1}(\sqrt{2} k_x) D_{j_1+1}(-\sqrt{2} k_x))
\eeqarr

However, one must note that in the presence of magnetic field, the
time reversal symmetry is broken, therefore, the other Dirac cone, the
time reversal partner, no long behaves the same way. So the dispersion
relation must be calculated separately. By the same approach, one can
get, for the $l.e.b.$,

\beqarr
&&\hspace{-0.5cm}(\epsilon ^2-j_1) D_{j_1} (-\sqrt{2} k_x) (j_2
D_{\epsilon ^2} (\sqrt{2} k_x)D_{j_{2-1}} (-\sqrt{2} k_x)\nonumber\\
&&+\epsilon ^2 D_{\epsilon  ^2-1} (\sqrt{2} k_x) D_{j_2} (-\sqrt{2} k_x)) \nonumber\\
&&\hspace{-0.5cm} = (\epsilon ^2-j_2) D_{j_2} (-\sqrt{2} k_x) (j_1
D_{\epsilon ^2} (\sqrt{2} k_x)D_{j_1-1} (-\sqrt{2} k_x)\nonumber\\
&&+\epsilon ^2 D_{\epsilon  ^2-1} (\sqrt{2} k_x) D_{j_1} (-\sqrt{2} k_x));
\eeqarr
for the $h.e.b.$,
\beqarr
&&\hspace{-0.5cm}(j_2+1)(\epsilon ^2-j_1-\gamma ^2)D_{j_1} (-\sqrt{2}
k_x)  (j_2 D_{\epsilon  ^2} (\sqrt{2} k_x) \nonumber\\
&&D_{j_{2-1}} (-\sqrt{2}k_x) +\epsilon ^2 D_{\epsilon ^2-1} (\sqrt{2} k_x)D_{j_2} (-\sqrt{2} k_x)) \nonumber\\
&&\hspace{-0.5cm}=(j_1+1) (\epsilon ^2-j_2-\gamma ^2) D_{j_2}
(-\sqrt{2}k_x)  (j_1 D_{\epsilon ^2} (\sqrt{2} k_x)\nonumber\\
&&\hspace{-0.5cm}D_{j_1-1}(-\sqrt{2} k_x)+\epsilon ^2 D_{\epsilon^2-1}
(\sqrt{2} k_x) D_{j_1} (-\sqrt{2} k_x)).
\eeqarr

 \begin{figure}
 \includegraphics[width=7cm]{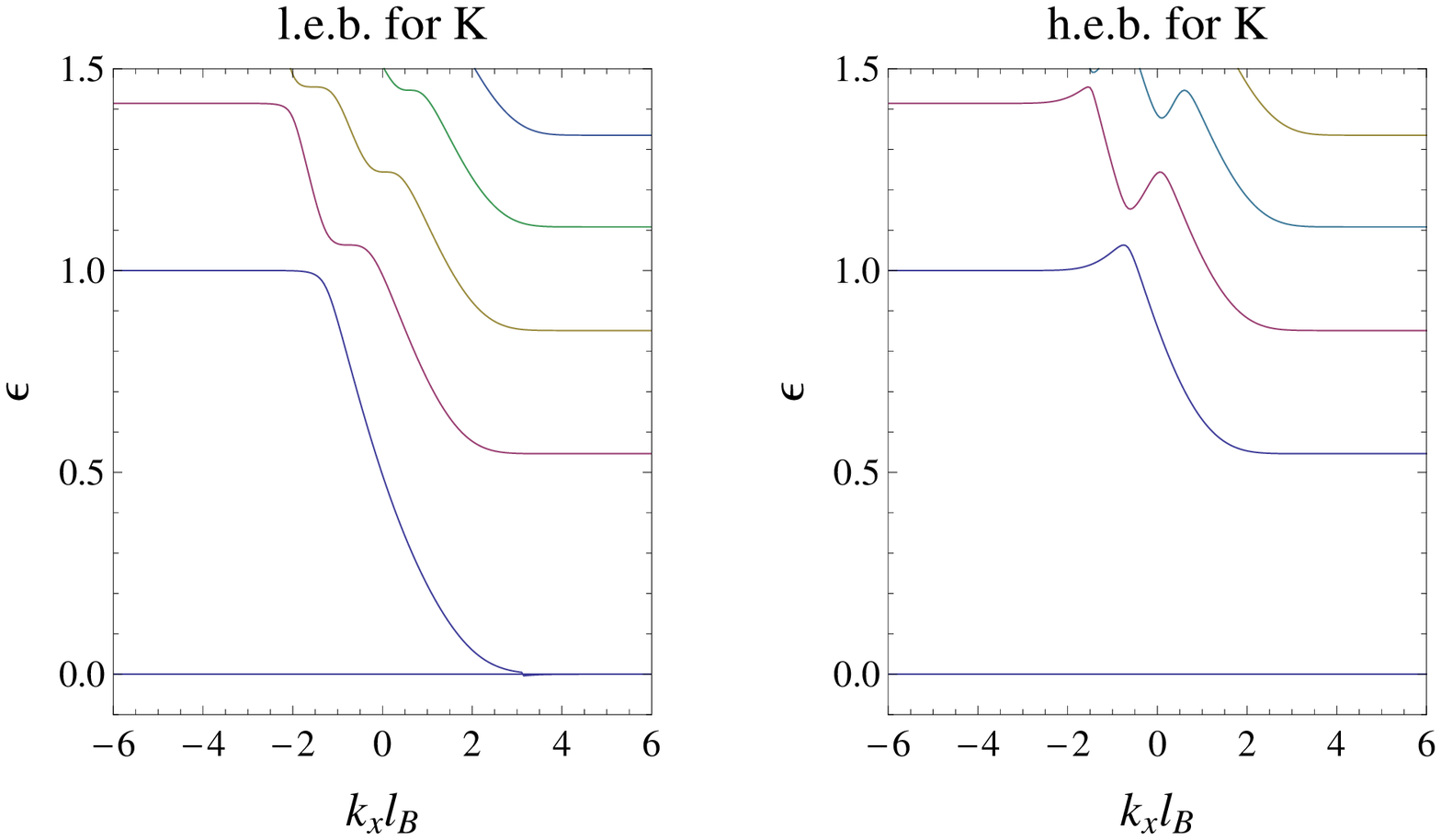}
  \includegraphics[width=7cm]{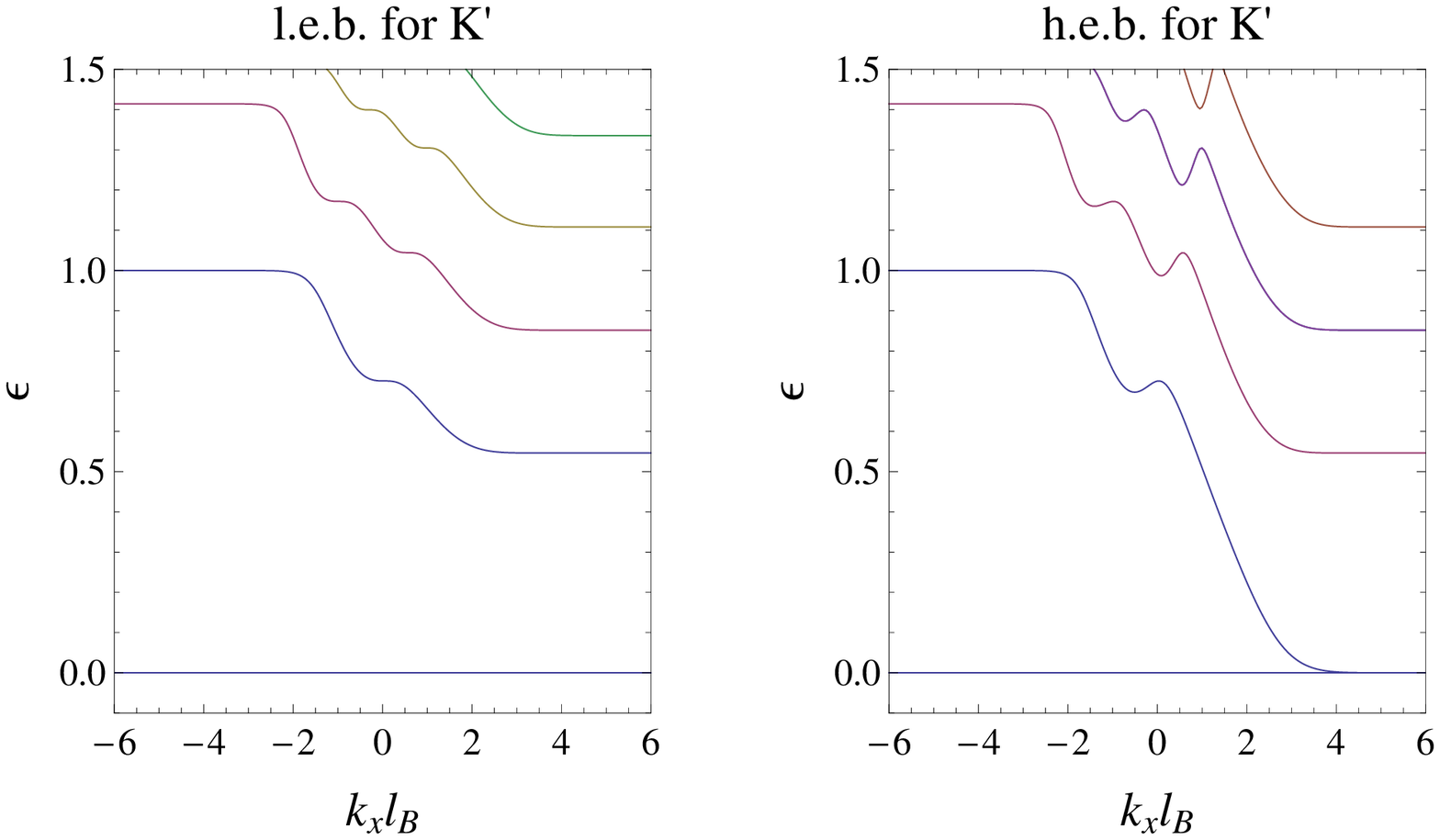}
 \caption{\label{disp:Bfield}(color online) The dispersion around Dirac cones in the
   presence of magnetic field. The upper two figures are for the {\it
     l.e.b.} and {\it h.e.b.} interfaces around one Dirac cone
   respectively, and  the lower two figures are that for the other
   Dirac cone.}
 \end{figure}

It is easy to see from the dispersion equation that when $\abs{k_x}$
is very large, the eigen-energy should restore to either the monolayer
value or the bilayer value as the parabolic cylinder functions $D_\nu$ only
converge to zero for integer $\nu$ at infinity. But what interests us
is how these bulk Landau levels connect with each other when crossing
the interface. To study that,we solve the
above equations for $\epsilon$ and $k_x$ near the interface $k_x =
0$ for a finite size system as shown in Fig.~\ref{disp:Bfield}. We see
that at the two Dirac cones, the Landau levels match in different ways.  
\begin{figure}
\includegraphics[width=6cm, height=5cm]{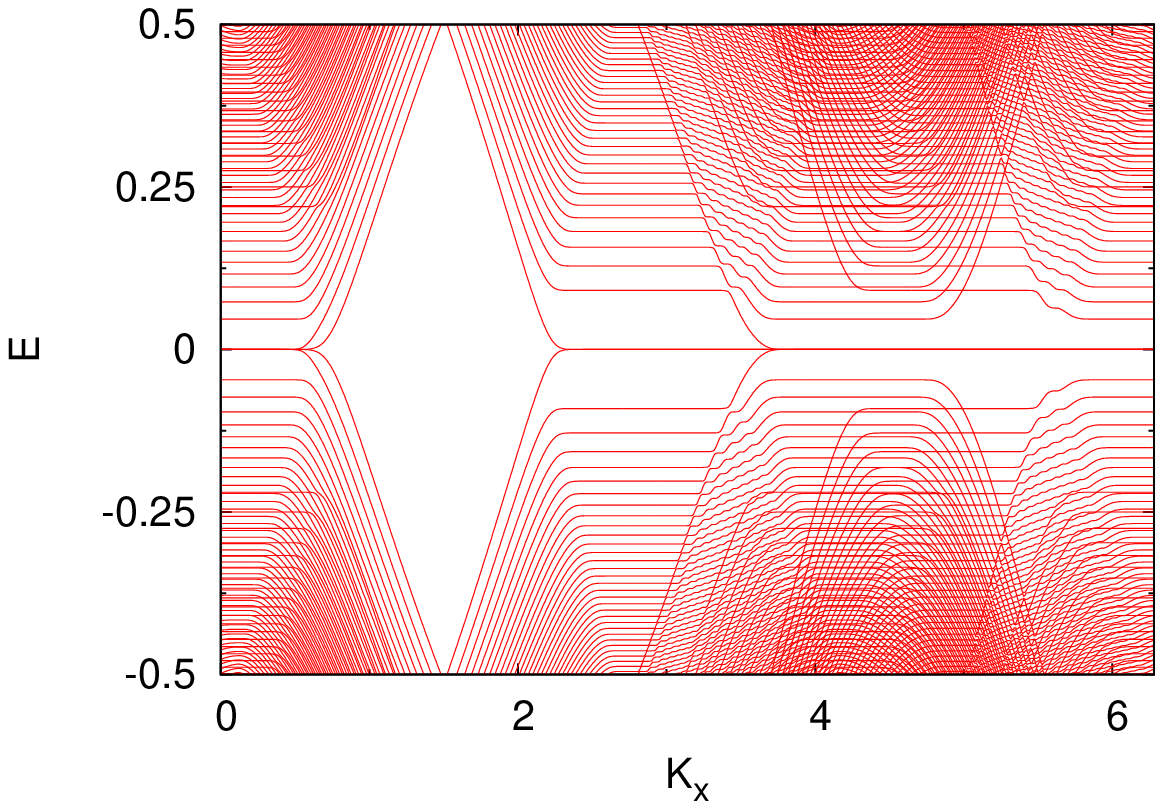}
\includegraphics[width=7cm]{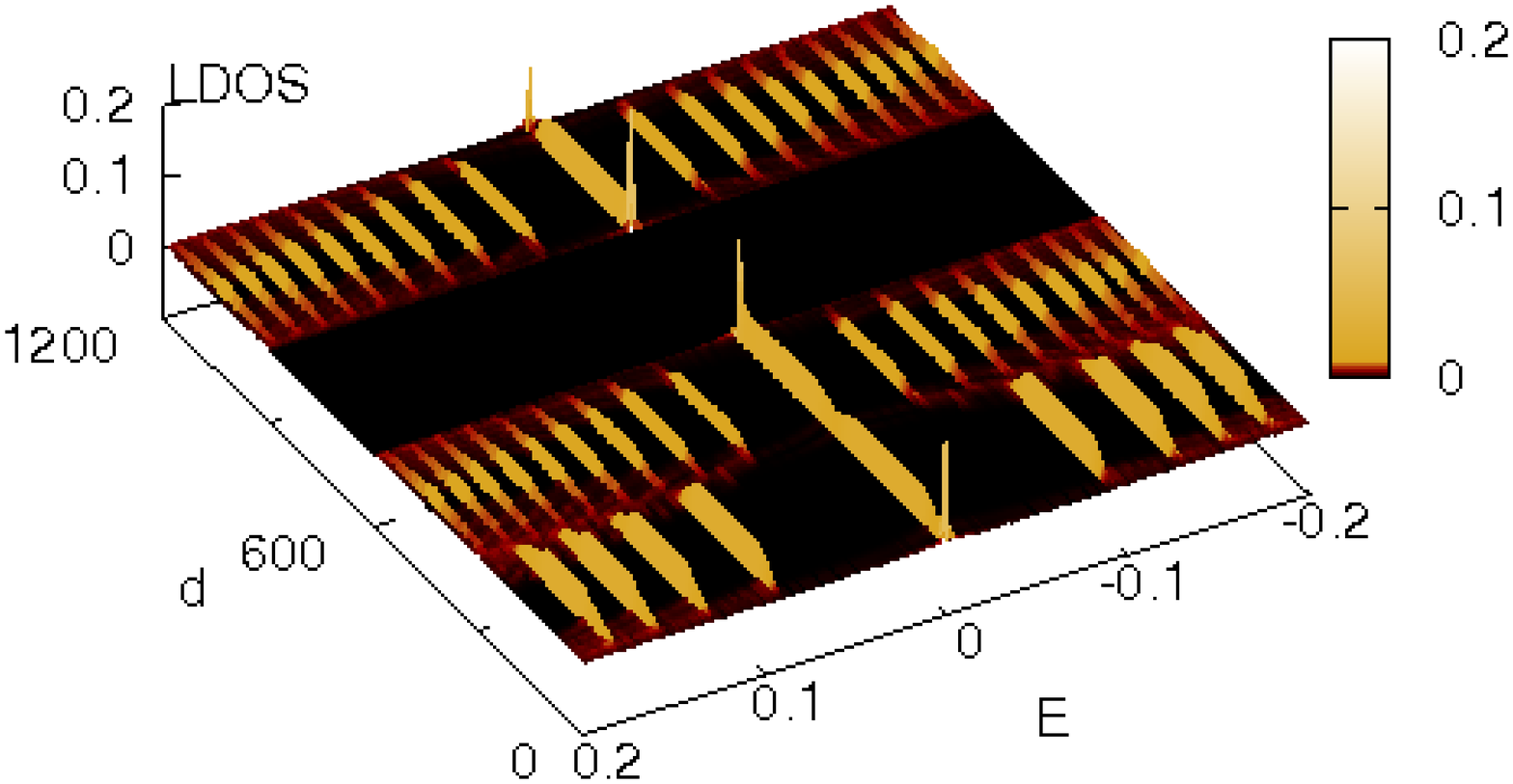}
\caption{\label{Blowenergy}The energy spectrum and LDOS of the
step bilayer graphene with {\it l.e.b.} in a
magnetic field. The strength of the magnetic field is expressed as
magnetic flux $\phi$ in each unit cell. Here we set the magnetic flux per each unit cell
$\phi=\phi_0/1315$ ($\phi_0$ is the magnetic flux quanta) which corresponds to $B \sim 60 T$.}
\end{figure}
\begin{figure}
\includegraphics[width=6cm, height=5cm]{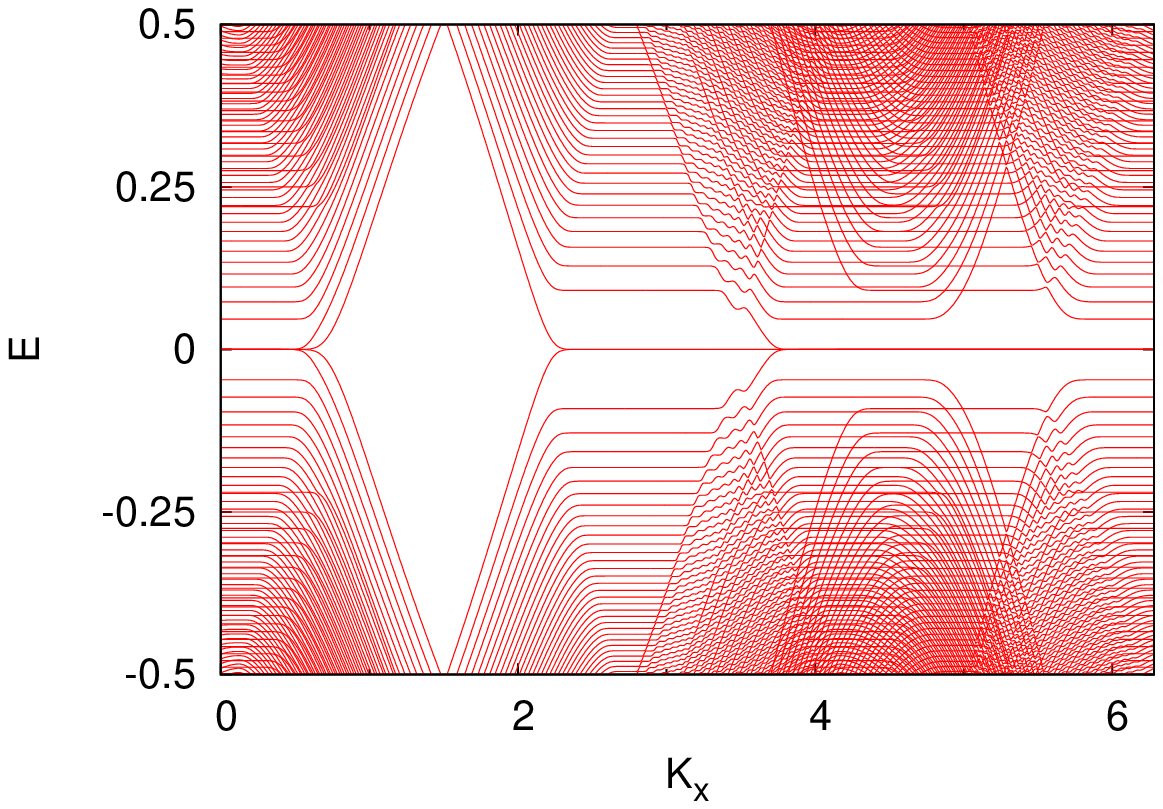}
\includegraphics[width=7cm]{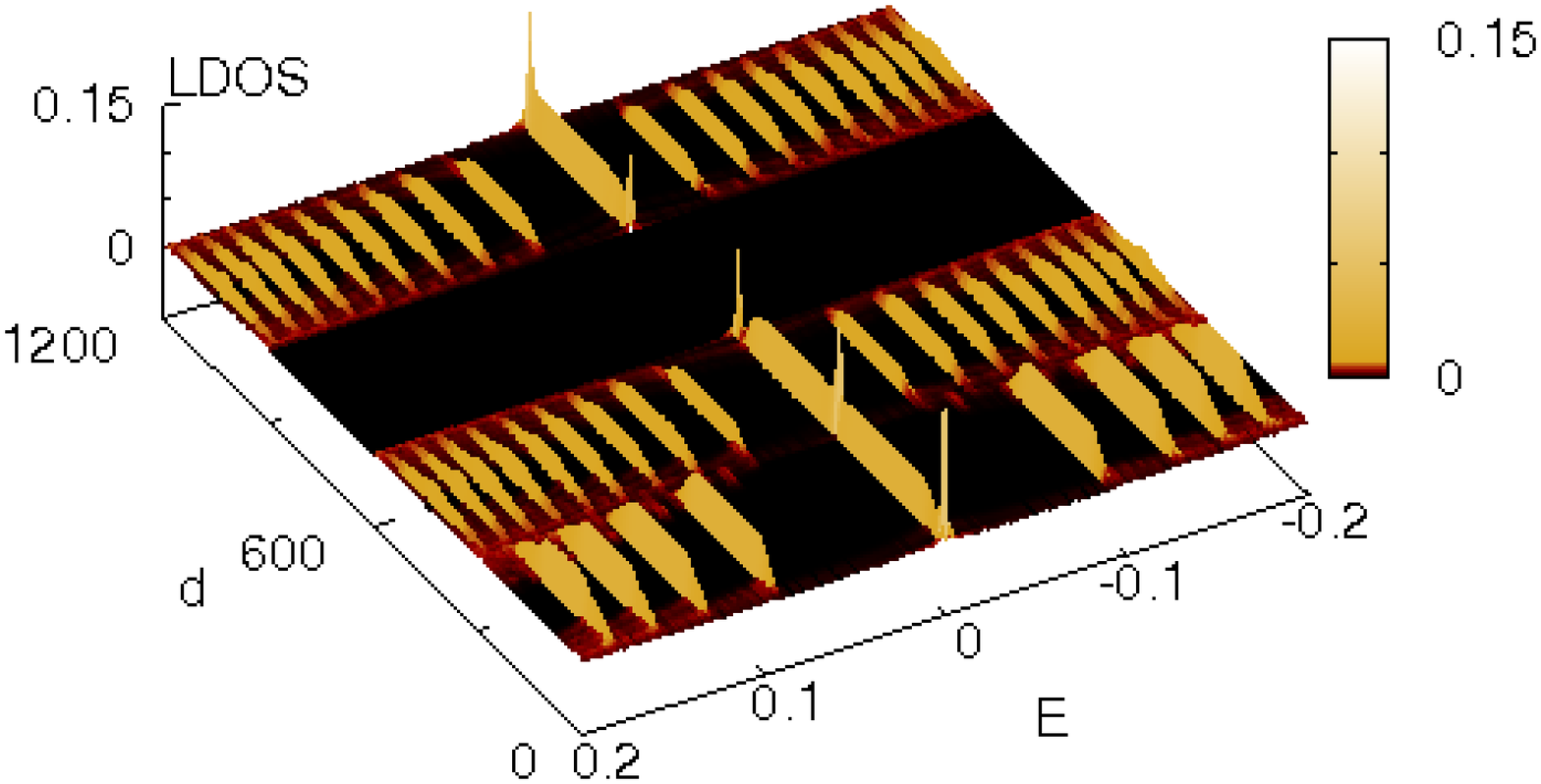}
\caption{\label{Bhighenergy}The energy spectrum and LDOS of the
step bilayer graphene with {\it h.e.b.} within magnetic field. The
parameters are the same as in Fig.~\ref{Blowenergy}}
\end{figure} 
Furthermore, we present the numerical result for directly diagonalizing a ribbon with width $L = 600$.
Our diagonalization results consistent with the analytic analysis also.
Fig.~\ref{Blowenergy} and Fig.~\ref{Bhighenergy} shows the energy
spectrum and LDOS for step bilayer graphene in a magnetic field
with {\it l.e.b.} and {\it h.e.b.} respectively. In magnetic
field the energy spectrum is split into a set of Landau levels(LLs)
and the Landau level splitting in bilayer is smaller than that of
monolayer graphene since the energy of LLs in bilayer system satisfies $E
\propto \sqrt{n(n-1)}$ and $E \propto \sqrt{n}$ in monolayer.
Therefore, LLs are not matched at the interface.

The major difference between the two sets of dispersion relation is
that the Landau levels in the two distinct regions connect in two
different ways. The dispersion at the interface with {\it l.e.b.} is more flat than that with
{\it h.e.b.}. These interface Landau levels was discussed by M. Koshino et.al.
within an effective-mass approximation~\cite{Koshino}. From the plot
of LDOS, the same as in the case of zero field, according to the different distribution of 
the edge state, the presence of the peak at $E=0$
along the interface only on the upper layer for {\it
  l.e.b.} and on both two layers for {\it h.e.b.}.

\begin{figure}
  \includegraphics[width=8cm]{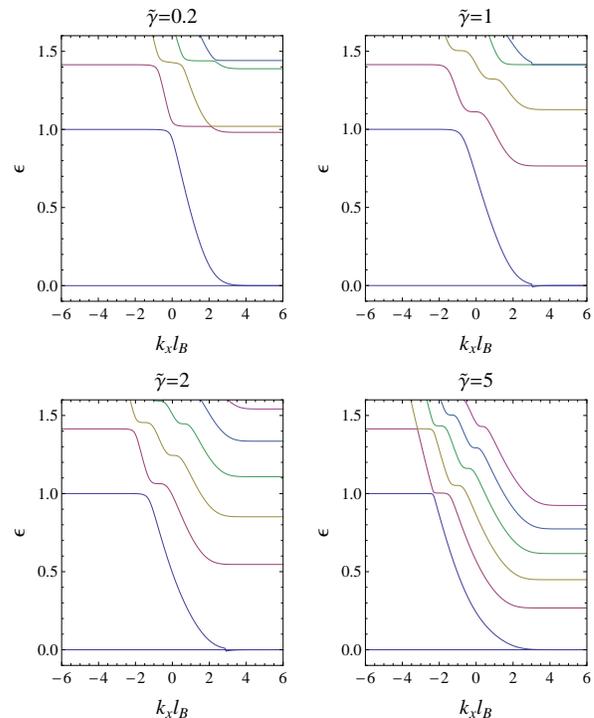}\\
   \caption{The dispersion around Dirac cones $K$ of $l.e.b.$ type
     interface in different magnetic field with $\tilde{\gamma} =
     0.2,1,2,5$ respectively.}\label{fig:disp_g1}
\end{figure}

\begin{figure}
    \includegraphics[width=8cm]{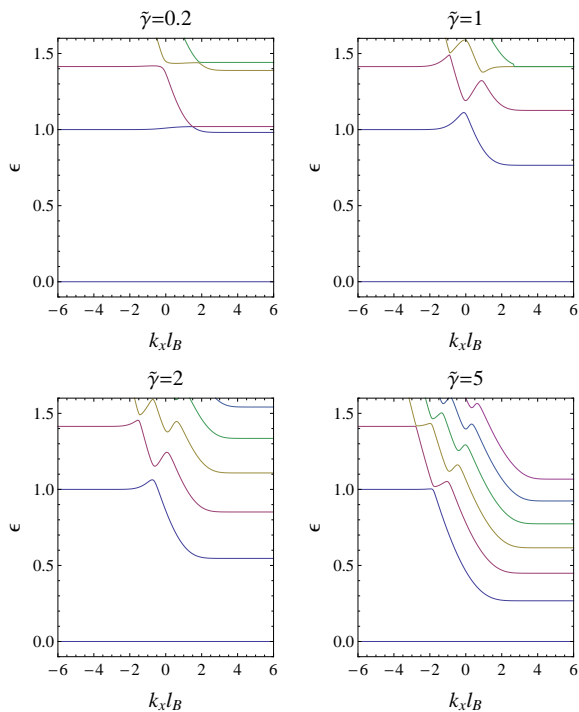}\\
   \caption{The dispersion around Dirac cones $K$ of $h.e.b.$ type
     interface in different magnetic field with $\tilde{\gamma} =
     0.2,1,2,5$ respectively.}\label{fig:disp_g2}
\end{figure}

We also use the Dirac equation result to study how the dispersive
Landau levels near the interface changes when the magnetic field
varies. If we measure the energy in unit of $\hbar \omega_c$, the
monolayer bulk Landau levels remain unchanged as the magnetic field is
being tuned; but for the bilayer, the effective interlayer coupling
$\tilde{\gamma} \sim 1/\omega_c$ so the bulk Landau levels vary with
the magnetic field. In strong field $\tilde{\gamma} \sim 0$, so the
two layers behave as if they were decoupled, and the Landau levels'
energy becomes just like the monolayer's but with a two-fold
degeneracy. In weak field limit, $\tilde{\gamma} \sim \infty$, in this
case one gets $\epsilon \simeq \sqrt{n(n-1)}/\tilde{\gamma}$. The
dispersions for $\tilde{\gamma} \in {0.2, 1,2,5}$ of both $l.e.b.$ and
$h.e.b.$ at one Dirac point are shown in
Fig.~\ref{fig:disp_g1} and Fig.~\ref{fig:disp_g2} respectively. Here we only plot the
Landau levels up to $n=6$. The evolution of dispersion on the other Dirac point have the similar 
behavior except different energy level connection as shown in Fig.~\ref{disp:Bfield}.
As the strength of the magnetic field is increased, the bilayer
LLs become more denser which induces more mediate states around the interface. The dispersion of the
$l.e.b.$ interface is always more flat than that of $h.e.b.$ interface while tuning the magnetic field.

\section{Discussion and Conclusion}
In this paper, we discuss the edge state at the interface between
monolayer and bilayer graphene. Physically, there are two types
of interface structures named by the $l.e.b.$ and $h.e.b.$ due to the different terminations at the interface. 
By studying the system with both effective theory and numerical calculation, we
find that with or without magnetic field, zero energy edge states
exist for both types of interface, however, the LDOS shows different
features. For $l.e.b.$ the zero energy edge states only live in the
upper layer which terminates at the interface while the $h.e.b.$
induces an enhanced LDOS in the bottom layer. Another major difference
between the two different geometries is that the dispersion of
$h.e.b.$ system shows a stronger anticrossing feature in the zero field. Both
differences can be interpreted as the following. When
an electron goes through the  $h.e.b.$ interface, it can choose to lower
its energy through the interlayer coupling. This can be viewed as the
$h.e.b.$ imposes an effective potential barrier at interface in the
extended layer. Since the localized zero energy edge states locates on both
two layers, the monolayer dispersion which is linear should smoothly connect 
to the edge state in the $h.e.b.$ case. This can explain the strong energy 
level anticrossing in the $h.e.b.$ interface. Similar differences between $l.e.b.$ and $h.e.b.$ interfaces are also
discussed for transimission coefficients across the
interface\cite{Nilsson,Takanishi,wenxin1}. For the $h.e.b.$ the
transmission probability is reduced significantly for incoming
electrons with energy $E \sim t_{\bot}$ which can be interpreted in a
similar way.

We also see that in presence of magnetic field the dispersion of Landau levels
continueously goes through the interface in different manners at the two Dirac
cones. At Dirac cone $K(K')$, the zero energy Landau levels in the bilayer
region split, one branch rises up and becomes the $n=1$ Landau levels
in the monolayer while the other branch remains as zero energy
states, and the other higher Landau levels continues from bilayer to
monolayer accordingly, i.e. $n=1 \big\vert_{\text{bi}}\rightarrow n
=2\big\vert_{\text{mono}}$ $\dots$. However, near the other Dirac
point $K'(K)$, the zero energy state remains intact, so the other higher Landau
levels connect as  $n=1 \big\vert_{\text{bi}}\rightarrow n
=1\big\vert_{\text{mono}}$ $\dots$. We also see an effect of the
effective potential barrier imposed by the $h.e.b.$ in the dispersive
Landau levels near the interface. For a certain range of field
strength the dispersion develops a local maximum for $h.e.b.$ while
the $l.e.b.$ only develops a plateau feature. 

In summary, the physical properties of hybrid graphene systems are
mostly dominated by both that of monolayer and bilayer. The Landau levels
remain the same away from the interface. But the existence of
dispersive Landau levels near the interface could be related to the
unexpected feature other than that of the monolayer and bilayer
graphene in the magneto transport experiment\cite{yliu}. Further study of the
hybrid structures for a more realistic setup, like including edge
disorder, gate voltage, etc., are needed to undertand the experimental
data.
\section{acknowlegement}
We wish to thank K. Yang and Y. Barlas for very helpful discussions.
This work is supported by Fundamental Research Funds for the Central Universities CDJRC10300007 (Z. X. H) and NSF under Grant No. DMR-1004545 (W.X.D.).

\end{document}